\documentclass{cupconf}
\usepackage{psfig}
\usepackage{amssymb}

  \checkfont{eurm10}
  \iffontfound
    \IfFileExists{upmath.sty}
      {\typeout{^^JFound AMS Euler Roman fonts on the system,
                   using the 'upmath' package.^^J}%
       \usepackage{upmath}}
      {\typeout{^^JFound AMS Euler Roman fonts on the system, but you
                   dont seem to have the}%
       \typeout{'upmath' package installed. cupconf.cls can take advantage
                 of these fonts,^^Jif you use 'upmath' package.^^J}%
      }
  \else
  \fi

  \checkfont{msam10}
  \iffontfound
    \IfFileExists{amssymb.sty}
      {\typeout{^^JFound AMS Symbol fonts on the system, using the
                'amssymb' package.^^J}%
       \usepackage{amssymb}%

      }{}
  \fi

  \IfFileExists{amsbsy.sty}
    {\typeout{^^JFound the 'amsbsy' package on the system, using it.^^J}%
     \usepackage{amsbsy}}
    {}

\newcommand{\ltsima}{$\; \buildrel < \over \sim \;$}
\newcommand{\simlt}{\lower.5ex\hbox{\ltsima}}
\newcommand{\gtsima}{$\; \buildrel > \over \sim \;$}
\newcommand{\simgt}{\lower.5ex\hbox{\gtsima}}

\newsavebox{\astrutbox}
\sbox{\astrutbox}{\rule[-5pt]{0pt}{20pt}}

\newcommand\mnras{Mon.~Not.~Roy.~Astr.~Soc.}
\newcommand\aj{Astron.~J.}
\newcommand\apj{Astrophys.~J.}
\newcommand\aaa{Astron.~Astrophys.}

\title[Galactic Structure]{Galactic Structure}

\author[Rosemary F.G.~Wyse]{R\ls O\ls S\ls E\ls M\ls A\ls R\ls Y\ls \ls \ls F.\ls G.\ls \ls \ls W\ls Y\ls S\ls E}
\affiliation{Department of Physics \& Astronomy, The Johns Hopkins University, Baltimore, MD 21218, USA}

\pubyear{2004}
\volume{xx}
\pagerange{xx}
\date{????}
\begin{document}

\maketitle

\begin{abstract}
Our Milky Way Galaxy is a typical large spiral galaxy, representative
of the most common morphological type in the local Universe.  We can
determine the properties of individual stars in unusual detail, and
use the characteristics of the stellar populations of the Galaxy as
templates in understanding more distant galaxies.  The star formation
history and merging history of the Galaxy is written in its stellar
populations; these reveal that the Galaxy has evolved rather quietly
over the last $\sim 10$~Gyr. More detailed simulations of galaxy
formation are needed, but this result apparently makes our Galaxy
unusual if $\Lambda$CDM is indeed the correct cosmological paradigm
for structure formation. While our Milky Way is only one galaxy, a
theory in which its properties are very anomalous most probably needs
to be revised. Happily, observational capabilities of next-generation
facilities should, in the the forseeable future, allow the aquisition
of detailed observations for all galaxies in the Local Group.
\end{abstract}

\section{Introduction: The Fossil Record}

The origins and evolution of galaxies, such as our own Milky Way, and
of their associated dark matter haloes are among the major outstanding
questions of astrophysics.  Detailed study of the zero-redshift
Universe provides complementary constraints on models of galaxy
formation to those obtained from direct study of high-redshift
objects.  Stars of mass similar to that of the Sun live for
essentially the present age of the Universe and nearby low-mass stars
can be used to trace conditions in the high-redshift Universe when
they formed, perhaps even the `First Light' that ended the
Cosmological Dark Ages.  While these stars may well not have formed in
the galaxy in which they now reside (especially if the CDM paradigm is
valid), several important observable quantities are largely conserved
over a star's lifetime -- these include surface chemical elemental
abundances (modulo effects associated with mass transfer in binaries)
and orbital angular momentum (modulo the effects of torques and
rapidly changing gravitational potentials).  Excavating the fossil
record of galaxy evolution from old stars nearby allows us to do
Cosmology locally, and is possible to some extent throughout the Local
Group, with the most detailed information available from the Milky Way
Galaxy. 

I here discuss our knowledge of the stellar populations of the Milky
Way and the implications for models of galaxy formation.
Complementary results for M31 are presented by Brown (this volume). 

\section{Large Scale Structure of the Stellar Components of the Galaxy}

There are four main stellar components of the Milky Way Galaxy.  

$\heartsuit$ The thin stellar disk: This is the most massive stellar component of the Milky Way, and  contains
stars of a wide range of ages and is the site of on-going star
formation.  A defining quality of the thin disk is that the stars are
on orbits of high angular momentum, close to circular orbits. The age
distribution of thin disk stars is not well determined even at the
solar circle, but there are clearly old, $\sim 10$~Gyr, thin disk
stars in the local vicinity.

$\heartsuit$ The thick stellar disk: This component, established some
20 years ago, has a scale-height around 3 times that of the thin
disk. Again its properties are best established close to the solar
Galactocentric distance; here the local thick disk consists of old stars,
that are on average of lower metallicity than that of a typical old star in
the thin disk, and are on orbits of lower angular momentum.

$\heartsuit$ The central bulge: This component is very centrally
concentrated and mildly triaxial with rotational energy close to the  expected value if
it were an oblate, isotropic rotator.  The dominant stellar population
is old and metal-rich. 

$\heartsuit$ The stellar halo: The bulk of the stars are old and metal-poor, 
on low angular-momentum orbits.  A few percent of the stellar mass is in globular clusters

\subsection{Large Scale Structure of the Thin Disk}

Our knowledge of the stellar populations in the thin disk is in fact
rather poor, with only very limited data on age distributions and
metallicity distributions in either the inner disk or the outer disk.  At
the solar neighbourhood, the gross characteristics of the metallicity
distribution of the thin disk have been known for a long time -- the
narrow distribution (peaking somewhat below the solar metallicity)
giving rise to the `G-dwarf Problem', or the deficit of metal-poor
stars compared to the predictions of the Simple closed-box model of
chemical evolution e.g.~van den Bergh 1962; Pagel \& Patchett 1975.
The favoured solution to this `problem' is to lift the `closed-box'
assumption, in particular to allow inflow of unenriched gas
(cf.~Larson 1972).  Such inflow is rather natural in many models of
disk formation and evolution (see Tosi's contribution to this volume). 

The age distribution of stars in the thin disk is particularly
important in setting the epoch of the onset of disk formation
(assuming that the bulk of the old stars now in the thin disk were
formed in the thin disk -- see Steinmetz's contribution to this volume
for an alternative view).  In Cold-Dark-Matter-dominated cosmologies,
the merging by which galaxies grow involves gravitational torques and
dynamical friction, which result in significant angular momentum
transport away from the central parts of dark matter haloes and their
associated galaxies, and into the outer parts.  This re-arrangement of
angular momentum is particularly effective if the merging involves
dense, non-dissipative (i.e.~stellar and/or dark matter) substructure
(cf.~Zurek, Quinn \& Salmon 1988; Zhang et al.~2002).

However, extended galactic disks as we observe them require detailed
angular momentum {\it conservation\/} during the dissipative collapse
and spin-up of proto-disk gas, within a dominant dark halo (cf.~Fall
\& Efstathiou 1980).  The angular momentum transport inherent in the
merging process in a CDM-universe results in disks that are too
concentrated and have radial scale-lengths that are too short
(cf.~Navarro, Frenk \& White 1995; Navarro \& Steinmetz 1997).  A
proposed solution to this problem delays the formation of stellar
disks until after the epoch of most merging, with `stellar feedback'
as the proposed mechanism for the delay (cf.~S.~Cole et al.~2000).  A
delay in radiative cooling -- the very first stage of stellar disk
formation -- until after a redshift of unity, or lookback times of
$\sim 8$~Gyr, is apparently required (Eke, Efstathiou \& Wright
2000). This has the obvious side-effect that the extended disks that
eventually form should contain few old stars.  Further, the
theoretical prediction is that disks form from the inside-out, with
lower angular momentum gas settling to the (inner regions of the) disk
plane earlier and only later settling of higher angular momentum material,
destined to form the outer disk.  The solar circle is some 2--3
exponential scale lengths from the Galactic center, and thus forms
later than the inner disk.  Detailed predictions are lacking (and
should be made), but allowing for 1~Gyr for cooling and star
formation, the expectation from this delay in disk formation would
then be that there should be very few stars in the local thin disk
that are older than $\sim 7$~Gyr.

Distances and kinematics derived for nearby stars from parallaxes and
proper motions from the Hipparcos satellite has allowed the
determination of the colour-absolute magnitude diagram for thin disk
stars.  Analysis of the locus in the CMD of subgiant stars gives a
lower limit to the age of the oldest stars of 8~Gyr, obtained with an
adopted upper limit to the metallicity of ${\rm [Fe/H] = +0.3}$
(Jimenez, Flynn \& Kotoneva 1998; Sandage, Lubin \& VandenBerg 2003);
the age limit increases approximately 1~Gyr for every 0.1~dex decrease
in the adopted metallicity (to set the age scale, the VandenBerg
isochrones used give ages of $\simgt 13$~Gyr for metal-poor globular
clusters).  Analysis of the main sequence turn-off stars in the
Hipparcos dataset provides a best-fit age for the oldest disk stars of
$\simgt 11$~Gyr (Binney, Dehnen \& Bertelli 2000), adopting a
metallicity distribution that peaks below the solar value and using
isochrones that provide an age of $\simgt 12$~Gyr for metal-poor
globular clusters.  It is clear that metallicity determinations for
the Hipparcos sample are needed before a definitive value for the age
of the oldest disk stars can be derived, but one should note that the
available metallicity distributions, mostly for G/K dwarfs, all peak
at ${\rm [Fe/H] = -0.2}$ (e.g.~Kotoneva et al.~2002), distinctly more
metal-poor than the +0.3~dex that gave the lower limit in age of 8~Gyr
for the oldest stars.  An older age is then expected. 

An alternative technique to derive ages uses the observed white dwarf (WD) 
luminosity function combined with theoretical models of white dwarf
cooling.  Hansen et al.~(2002; see also Richer's contribution to this
volume) analysed the disk WD luminosity function of Leggett
et al.~(1998) together with their own data for the WD 
sequence in the globular cluster M4.  They derived a $\sim 5$~Gyr gap
between the formation of M4 and the birth of the oldest disk stars,
with ages of $\sim 13$~Gyr and $\sim 8$~Gyr respectively.  However,
completeness remains an issue for the disk WD luminosity
function, and different determinations are available and provide older
ages and less of a gap in age (e.g.~Knox, Hawkins \& Hambly
1999). Indeed, a recent re-analysis of the M4 data (De Marchi et
al.~2003) has demonstrated that with a re-assessment of the errors,
the derived WD luminosity function is in fact still rising at the last
point and so only a lower limit in age, $\simgt 8$~Gyr, can be
derived. Clearly more and better data are needed, and should be
available, from the ACS on HST for M4, and from  surveys such
as SDSS for the faint, local disk WDs.

The star formation history of the local disk that is derived from 
the Hipparcos CMD (Hernandez, Valls-Gabaud \& Gilmore 2000)  has an amplitude that
shows a slow overall decline, with quasi-periodic increases of  a factor
of a few on timescales of $\sim 1$~Gyr.  This result is consistent with 
other age indicators such as chromospheric activity (Rocha-Pinto et
al.~2000), and with chemical evolution models. 

The available data are then all consistent with a significant
population of stars in the local disk with ages $\sim 8$~Gyr, and
perhaps as old as 11~Gyr.  If these stars formed in the disk, then the
formation of extended disks was {\it not\/} delayed until after a
redshift of unity, as was proposed to `solve' the disk angular momentum
problem in CDM models. 

The outer disk of M31 also contains old stars (Ferguson \& Johnson
2001; Guhathakurta this volume) and similar conclusions hold.
Further, deep, high-resolution IR observations have revealed
apparently relaxed disk galaxies at $z \simgt 1$ (Dickinson 2000),
which presumably formed at least a few dynamical times
earlier. Indeed, a candidate old disk has been identified at $z \sim
2.5$ (Stockton et al.~2003). 

\subsection{Large Scale Structure of the Thick Disk}

The thick disk was defined through star counts 20 years ago (Gilmore
\& Reid 1983) and is now well-established as a distinct component.
Its origins remain the source of considerable debate.  Locally, some
$\sim 5\%$ of stars are in the thick disk; the vertical scale-height is $\sim
1$~kpc, and radial scale-length $\sim 3$~kpc.  Assuming a smooth
double-exponential spatial distribution with these parameter values, the stellar mass of the
thick disk is 10--20\%. of that of the thin disk (the uncertainty
allowing for the uncertainty in the structural parameters), or some
$10^{10}$M$_\odot$. 

Again the properties of the stellar populations in this component are
rather poorly known far from the solar neighborhood. Locally, within a
few kpc of the Sun, the typical thick disk star is of intermediate
metallicity, ${\rm [Fe/H] \sim -0.6}$~dex, and old, with an age
comparable to that of 47~Tuc, the globular cluster of the same
metallicity, $\sim 12$~Gyr (see e.g.~review of Wyse 2000).  Detailed elemental
abundances are now available for statistically significant sample
sizes. These show that the pattern of elemental abundances differs
between the thick and thin disks, with different values of the ratio
[$\alpha$/Fe] at fixed [Fe/H], implying distinct star formation and
enrichment histories for the thick and thin disks (Fuhrmann 1998, 2000; Prochaska et al.~2000; Feltzing,
Bensby \& Lundstr\"om 2003; Nissen 2003; see Figure~1).  Such a difference argues against the
model (Burkert, Truran \& Hensler 1992) whereby the thick disk
represents the earliest stages of disk star formation during
continuous, self-regulated dissipational settling of gas to the thin
disk.

\begin{center}
\begin{figure}%[ht!]
\psfig{figure=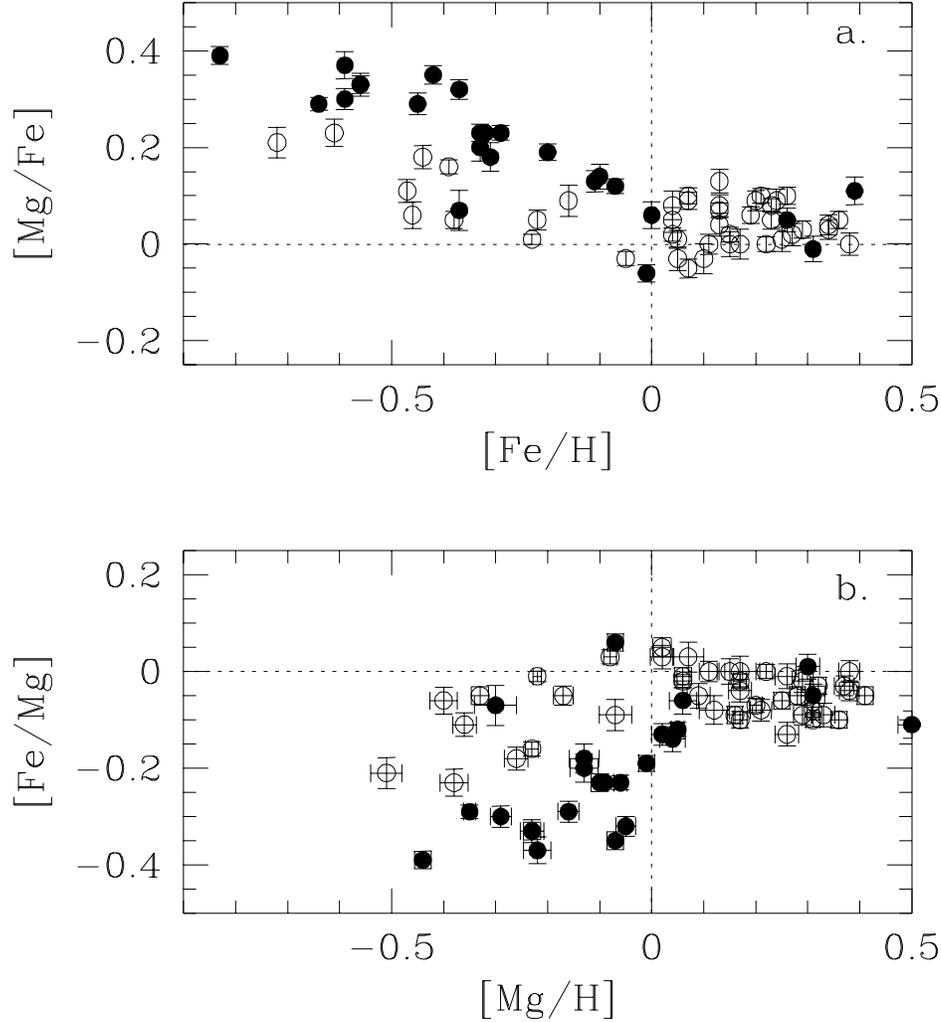}%,height=5in,width=5.2in}%,angle=270}
\caption{Taken from Feltzing et al.~2003, their Figure~2. Filled
symbols represent stars whose kinematics are consistent with
membership of the thick disk, while open symbols represent thin disk
stars.  The uncertainties in Mg abundance are indicated by the error
bars; uncertainties in Fe are smaller than the symbol sizes.  At a
given value of [Fe/H], the thick and thin disk stars are separated,
with thick disk stars having higher [Mg/Fe].  At the typical thick
disk metallicity, [Fe/H]$ \sim -0.5$~dex, the value of [Mg/Fe] in thick disk stars is
equal to that seen in the stellar halo, and consistent with enrichment by Type II supernovae.  More metal-rich thick disk stars show some enrichment by iron-dominated ejecta from Type Ia supernovae.}
\end{figure}
\end{center}

Thick stellar disks can be formed from pre-existing thin stellar disks
by heating, and a (minor) merger of a reasonably dense and massive
satellite galaxy into a pre-existing thin disk galaxy could be the
heating mechanism (cf.~Quinn, Hernquist \& Fullagar 1993).  In the
merger, orbital energy is deposited in the internal degrees of freedom
of both the thin disk and the satellite, and acts to disrupt the
satellite and heat the disk.  Depending on the orbit of the satallite, and on its density profile
and mass (this last determines the dynamical friction timescale),
tidal debris from the satellite will be
distributed through the larger galaxy during the merger process.  
Thus the phase space structure
of the debris from the satellite depends on many parameters, but in
general one expects that the final `thick disk' will be a mix of
heated thin disk and satellite debris.  The age and metallicity
distributions of the thick disk can provide constraints on the mix.

Could the thick disk be dominated by the debris of tidally disrupted
dwarf galaxies (cf.~Abadi et al.~2003)?  The removal of material from
the satellite occurs under essentially the Roche relative density
criterion, so that one expects that the lower density, outer parts of
accreted dwarfs will be tidally removed in the outer parts of the
larger galaxy, with the inner, denser regions of the dwarf only being
removed if the dwarf penetrates further inside the larger galaxy.
 As noted above, the local (within a few kpc
of the Sun) thick disk is old and quite metal-rich, with a mean iron
abundance $\sim -0.6$~dex.  Further, the bulk of these stars have
enhanced, super-solar [$\alpha$/Fe] abundances (Fuhrmann 1998, 2000;
Prochaska et al.~2000; Feltzing et al.~2003; see Figure~1).  Achieving such a high
level of enrichment so long ago (the stellar age equals the age of
47~Tuc, at least 10~Gyr, as noted above), in a relatively short time
-- so that Type II supernovae dominate the enrichment, as evidenced by
the enhanced levels of [$\alpha$/Fe]) -- implies a high star formation
rate within a rather deep overall potential well.  This does not favor
dwarf galaxies.

Indeed, the inner disk of the LMC, our present most massive satellite
galaxy, has a derived metallicity distribution (Cole, Smecker-Hane \&
Gallagher 2000) that is similar to that of the (local) thick disk, but, based on the color-magnitude diagram, 
these stars are of intermediate age. Thus the LMC apparently took until a
few Gyr ago to self-enrich to an overall metallicity that equals that of
the typical local thick disk star in the Galaxy.  Further, the
abundances of the $\alpha$-elements to iron in such  metal-rich LMC
stars are below the solar ratio (Smith et al.~2002), unlike the local thick disk stars.  This may be understood in terms of the different star-formation histories (cf.~Gilmore \& Wyse 1991).  The LMC is not a
good template for a putative dwarf to form the thick disk from its debris. 

What about the Sagittarius dwarf, a galaxy that has clearly penetrated
into the disk?  From photometry the stars are on average quite
enriched and of intermediate age (cf.~the discovery paper of Ibata,
Gilmore \& Irwin 1994, where the member stars were clearly
distinguished from the bulge field stars; see also Layden \&
Saradejini 2000 and Cole 2001). The overall metallicity distribution of the
Sagittarius dwarf spheroidal galaxy is not well defined, but it
contains a significant population of stars with metallicity as high as the solar value (Bonifacio et al.~2000; Smecker-Hane \&
McWilliam 2003).  One of its globular clusters, Terzian~7, has a
metallicity equal to that of 47~Tuc, but an age several Gyr younger
(Buonannno et al.~1995), and thus by inference, several Gyr younger
than the thick disk stars of the same metallicity.  The derived
age-metallicity relationship for the Sgr dSph, based on both the CMD
(Saradejini \& Layden 2000) and spectroscopy of selected red giants
(Smecker-Hane \& McWilliam 2003), is consistent with stars more
metal-rich than [Fe/H] $=-0.7$~dex being less than 8~Gyr old.
Further, these stars have essentially solar values of the ratio of
[$\alpha$/Fe] (Bonifacio et al.~2000; Smecker-Hane \& McWilliam 2003), and are thus different in several important properties 
from  the local thick disk stars.  The Ursa Minor dSph is the only satellite
galaxy of the present retinue that contains only old stars, and thus has an age distribution similar to that of the local thick disk.  However
these are exclusively metal-poor, ${\rm [Fe/H] \sim -2}$~dex, and
again not a good match to  thick disk stars.

Thus based on observations, there is no good analogue among the
surviving dwarf galaxies for a possible progenitor of the thick disk.
Theoretically, based on our (admittedly limited) understanding of
supernova feedback, it seems very contrived to envisage a dwarf galaxy
that had a deep enough potential well to self-enrich rapidly a long
time ago, but that was sufficiently low density to be tidally
disrupted to form the thick disk.  One might argue that the satellites
that were accreted earlier, initiated star formation earlier
(e.g.~Bullock, Kravtsov \& Weinberg 2000) and were typically more
dense and able to self-enrich faster. However, the analyses of deep
color-magnitude diagrams for the extant satellites of the Milky Way
are consistent with {\it all\/} containing stars as old as the stellar
halo of the Milky Way (e.g.~Da Costa 1999), implying that the onset of
star formation was co-eval and there are no (surviving) satellites
that initiated star formation earlier.

In summary, it appears 
implausible that the bulk of the thick disk is the debris of
accreted dwarf galaxies.

Heating of a pre-existing thin disk by a minor merger remains a viable
mechanism for creating the bulk of the thick disk (e.g.~Velazquez \& White
1999). In this case, the old age of the thick disk, combined with the
fairly continuous star formation in the thin disk, has two important
consequences -- the first that there was an extended disk in place at
a lookback time of greater than 10~Gyr, and the second that there has
been no extraordinary heating -- by mergers -- of the thin disk since
that time.  Knowing the age distribution of stars in the thick disk --
in both the observed thick disk and in predicted theoretical thick
disks -- is obviously crucial. Semi-analytic
modelling of the heating of disks by merging of substructure in CDM
cosmologies has shown that thin disks with reasonable scale-heights
are produced at the present day (Benson et al.~2003).  However the
presence or otherwise of thick disks has yet to be demonstrated in such simulations. Further, predictions need to be made 
for the age distribution of member stars of the
thick and the thin disk, to be confronted with the observations.

All this being said, some fraction of the metal-poor stars assigned to
the `thick disk', on the basis of having orbital kinematics that are
intermediate between those of the stellar halo and those of the thin
disk, may well be debris from a satellite (the one that caused the disk heating perhaps),
and we return to this point below, in section 3.2.

\subsection{Large Scale Structure of the Central Bulge}

The metallicity distributions of low-mass stars in various
low-reddening lines-of-sight towards the bulge (with projected
Galactocentric distances of a few 100pc to a few kpc) have been
determined spectroscopically (e.g. McWilliam \& Rich 1994; Ibata \&
Gilmore 1995; Sadler, Terndrup \& Rich 1996) and photometrically
(e.g.~Zoccali et al.~2003) with the robust result that the peak
metallicity is ${\rm [Fe/H]} \sim -0.3$~dex, with a broad range and a
tail to low abundances (indeed the distribution is well-fit by the
Simple closed-box model, unlike the solar neighborhood data).  The
available elemental abundances, limited to the brighter stars, show
the enhanced ${\rm [\alpha/Fe]}$ signatures of enrichment by
predominantly Type II supernovae (McWilliam \& Rich 1994; McWilliam 
\& Rich  2003), indicating rapid star formation.  Indeed
the chemical abundances favor very rapid star formation and (self-)enrichment
(Ferreras, Wyse \& Silk 2003).  

The age distributions derived from the
analyses of deep HST and ISO color-magnitude diagrams -- again over
several degrees across the sky -- are consistent with the dominant population being
of old age $\simgt 10$~Gyr (Feltzing \& Gilmore 2000 (HST); van Loon
et al.~2003 (ISO); Zoccali et al.~2003 (HST)), confirming earlier
conclusions  from ground-based data (Ortolani et al.~1995).  There is
also a small intermediate-age component seen in the ISO data, and
traced by OH/IR stars (Sevenster 1999), plus there is ongoing star
formation in the plane.  The interpretation of these younger stars in
terms of the stellar populations in the bulge is complicated by the
fact that the scale-height of the thin disk is comparable to that of
the central bulge, so that membership in either component is
ambiguous.  Indeed the relation between the inner triaxial bulge/bar
and the larger-scale bulge is as yet unclear (see Merrifield 2003 for
a recent review).  All that said, the dominant population in the bulge
is clearly old and metal-rich.

In the hierarchical clustering scenario, bulges are built up during
mergers, with several mechanism contributing.  The dense central
regions of massive satellites may survive and sink to the center; the
dynamical friction timescale for a satellite of mass $M_{sat}$
orbiting in a more massive galaxy of mass $M_{gal}$ is $t_{dyn\, fric}
\sim t_{cross} M_{gal}/M_{sat}$, where $t_{cross} $ is the crossing
time of the more massive galaxy.  With $t_{cross} \sim 3 \times 10^8$~yr for a large galaxy,  only the most massive
satellites could contribute to the central bulge in a Hubble time.  Gravitational torques
during the merger process are also expected to drive disk gas to the
central regions, and some fraction of stars in the disk will also be heated sufficiently to be 
`re-arranged' into a bulge (cf.~Kauffmann 1996).  The predicted age and
metallicity distributions of the stars in the bulge are then 
dependent on the merger history; however a uniform old
population is not expected. 

An alternative scenario for bulge formation appeals to an instability
in the disk, forming first a bar which then buckles out of the plane
to form a bulge (e.g.~Raha et al.~1991) or is destroyed by the
orbit-scattering effects of the accumulation of mass at its center
(e.g.~Hasan \& Norman 1990).  Again one would expect a significant
range of stellar ages in the bulge.

As noted above, the bulge is dominated by old, metal-rich stars.  This
favors neither of the two scenarios above, but rather points to
formation of the bulge by an intense burst of star formation, {\it in
situ}, a long time ago (cf.~Elmegreen 1999; Ferreras, Wyse \& Silk
2003). The inferred star formation rate is $\simgt 10M_\odot$/yr. A
possible source of the gas is ejecta from star-forming regions in the
halo; the rotation of the bulge is consistent with collapse
and spin-up of halo material (cf.~Wyse \& Gilmore 1992; Ibata \& Gilmore 1995), and the chemical abundances are also consistent with the mass ratios (see Carney, Lathm \& Laird 1990 and Wyse \& Gilmore 1992).

\subsection{Large Scale Structure of the Stellar Halo}

The total stellar mass of the halo is $\sim 2 \times 10^9M_\odot$
(cf.~Carney, Laird \& Latham 1990), modulo uncertainties in the
stellar halo density profile in each of the outer halo, where
substructure may dominate, and the central regions, where the bulge
dominates.  Some $\sim 30\%$ of the stars in the halo are on orbits
that take them through the solar neighborhood, to be identified by
their `high-velocity' with respect to the Sun.  These stars form a
rather uniform population -- old and metal-poor, with enhanced values
of the elemental abundance ratio [$\alpha$/Fe].  The dominant
signature of enrichment by Type~II supernovae indicates a short
duration of star formation.  This could naturally arise due to star
formation and self-enrichment occuring in low-mass star-forming
regions that cannot sustain extended star formation.  

In contrast, the typical star in a dwarf satellite galaxy now is of
intermediate-age, and has solar values of [$\alpha$/Fe] (cf.~Tolstoy
et al.~2003).  These differences in stellar populations between the
field stellar halo and dwarf galaxies limit significant ($\simgt 10$\%
by mass) accretion into the stellar halo from satellite galaxies to
have occurred at high redshift only, at lookback times greater than
$\sim 8$~Gyr (cf.~Unavane, Wyse \& Gilmore 1996).  Typical CDM-models
predict, in contrast, significant late accretion of sub-haloes, with
around 40\% of subhaloes that survive reionization falling into the
host galaxy at redshifts less that $z=0.5$, or a look-back time of
less than 6~Gyr (Bullock, Kravtsov \& Weinberg 2000).  Again the
later accretion is preferentially to the outer parts, and to be consistent
with the observations of the Milky Way, these sub-haloes must contain
very few young stars, and not over-populate the outer galaxy with
visible stars. 

\subsection {Large Scale Structure: Merging History}

The overall properties of the main stellar components of the Milky
Way, as discussed above, can be understood if there was little merging
or accretion of stars into the Milky Way for the last$\sim 10$~Gyr (cf.~Wyse 2001).
How does this compare with `merger trees' of N-body simulations?  As
an example, the publically available Virgo GIF $\Lambda$CDM
simulations (Jenkins et al.~1998) have 26 final haloes with mass similar
to that of the Milky Way -- taken to be $2 \times
10^{12}$~M$_\odot$. Of these, only 7\% have not merged with another
halo of at least 20\% by mass since a redshift of 2, a look-back time
of $\sim 11$~Gyr in this cosmology (L.~Hebb, priv.~comm.).  A merger with these parameter
values could produce a thick disk as observed in the Milky Way.  None
of these `Milky Way' analogues pass a more stringent mass ratio limit
of no mergers more than 10\% by mass (still capable of producing a
thick disk, given appropriate orbit etc.) since a redshift of 2.  
Reducing the epoch of last significant merger to unity (a lookback
time of 8~Gyr) and adopting a maximum  merging mass
ratio of 20\%, makes the Milky Way more typical, with 35\% of Milky Way analogues meeting these
criteria. However reducing the highest mass ratio to 10\%, while
maintaining this lower look-back time limit, results in only 4\% of
Milky Way analogue haloes passing these criteria.  The
Milky Way appears to be rather unusual in the $\Lambda$CDM cosmology.

Note that predictions of smooth average `universal' mass assembly histories (e.g.~Wechsler et
al.~2002) are not useful for this comparison, since these curves suppress the 
detailed information necessary to predict the effect of the mass accretion.   More useful is the detailed merging history as a function
of radius (cf.~Helmi et al.~2003a), since ideally one would like to
know what fraction of mergers can penetrate into the realm of the baryonic  Galaxy.

%\subsection{Dark Halo  Large Scale Structure}

%something here?? Didn't talk about it ...put in intro?

\section{The Small Scale Structure of the Stellar Components of the  Galaxy}

While there is no evidence for recent very significant mergers into
the Milky Way, mergers are clearly happening, as best evidenced by the
Sagittarius dwarf galaxy (Ibata, Gilmore \& Irwin 1994, 1995; Ibata et
al.~1997; see Majewski's contribution to this volume).  While the present and past mass of the Sagittarius dSph
are rather uncertain, its assimilation into the Milky Way is best classed as a `minor
merger', meaning mass ratio of less than 10\%.

The small-scale structure in the Milky Way may reflect the
minor-merger history -- or may simply reflect inhomogeneities of
different kinds.

\subsection{Small Scale Structure in the Thin Disk }

The small scale structure of the thin disk is rich and varied and
includes stellar moving groups, the scatter in the age-metallicity
relationship, spiral arms, the outer `ring' structure and the central
bar.  All star formation appears to occur in clusters (e.g.~Elmegreen
2002), which are then subject to both internally and externally driven
dynamical processes that operate to disrupt them. Some clusters
dissolve almost immediately star formation is initiated and some
remain gravitationally bound for many Gyr. The creation of phase space
structure is thus a natural part of the evolution of stellar disks. 

The scatter in the age-metallicity relationship for stars at the solar
neighborhood appears to be well-established (Edvardsson et al.~1993),
as is the offset between the metallicity of the Sun, and of younger
stars and the interstellar medium, in the solar neighbourhood, with
the Sun being more chemically enriched.  These may have their origins
in some combination of radial gradients and mixing (e.g.~Francois \&
Matteucci 1993; Sellwood \& Binney 2002) and infall of metal-poor gas,
from the general intergalactic medium, or perhaps from satellite
galaxies (e.g.~Geiss et al.~2002).  It should be noted that further
motivation for appeal to accretion events from satellite galaxies had
been found in the scatter in element ratios at a given iron abundance
in the Edvardsson et al.~data.  However,  more recent data has instead 
suggested that
a more correct interpretation is that the elemental abundances of
stars belonging to the thick disk are distinct from those belonging to
the thin disk (e.g.~Nissen 2003; see also Gilmore, Wyse \& Kuijken
1989 and Figure~1 above), with no scatter in elemental ratios within a given component. 

The low-latitude `ring' seen in star counts (Newberg et al.~2002;
Ibata et al.~2003; Bellazzini et al.~2004)  in the anticenter
direction at Galactocentric distances of $\sim 15$~kpc may be
structure in the outer disk, which has a well-established warp in the
gas, and probably in stars (e.g.~Carney \& Seitzer 1993; Djorgovski \&
Sosin 1989).  The recent detection of structure in HI, interpreted as
a newly identified spiral arm, at just this distance
(McClure-Griffiths et al.~2003) is intriguing.  The ring may also be
interpreted as resulting from the accretion of a satellite galaxy
(e.g.~Helmi et al.~2003b; Martin et al.~2004; Rocha-Pinto et
al.~2003); a recent N-body hydrodynamic simulation within a
$\Lambda$CDM cosmology has shown that it is possible for satellite
galaxies to be accreted into a disk, provided they are massive enough
for dynamical friction to circularize their orbit quickly enough
(Abadi et al.~2003).

The available kinematics for `ring' stars do not discriminate between
the two possibilities of satellite or outer disk (Yanny et al.~2003;
Crane et al.~2003).  Given the complexity of the structure of outer
disks, comprehensive color-magnitude data plus metallicity
distributions plus kinematics will be needed to rule out the `Occam's Razor' 
interpretation of the `ring' as being a manifestation of structure in
the outer disk.

The detailed structure of the thin disk will be revealed by
large-scale spectroscopic surveys such as RAVE (Steinmetz 2003); the time is ripe to develop models that will 
distinguish between intrinsic structure due to the normal disk 
star formation process and other effects (cf.~Freeman \& Bland-Hawthorn 2002).

\subsection{Small Scale Structure in the Thick Disk }

As noted above, in the minor-merger scenario for formation of the
thick disk, one expects the `thick disk' to be a mixture of heated
thin disk, plus satellite debris.  An identification of satellite
debris, made on the basis of distinct kinematics, was made by Gilmore,
Wyse \& Norris (2002). These authors obtained radial velocities for
several thousand faint ($V \simlt 19.5$) F/G dwarf stars, selected by
photometry to be unevolved stars in the thick disk/halo interface at
several kpc from the Sun, in key intermediate-latitude lines of sight
that probe orbital rotational velocity (particularly $\ell =
270^\circ$).  They found that the mean lag behind the Sun's azimuthal
streaming velocity was significantly larger for the fainter stars that
for the brighter stars (see Figure~2).  Either there are (discontinuous?) steep kinematic 
gradients within the thick disk (cf.~Majewski 1993), or a separate
population exists. In the latter case, a viable explanation would be
stars from a shredded satellite.
Indeed, these stars have low metallicities, typically $-1.5$~dex
(Norris et al., in prep), a factor of ten below a typical thick disk
star (e.g.~Gilmore, Wyse \& Jones 1995), but typical of the old
population in dwarf galaxies. 

\begin{figure}%[ht!]
\psfig{figure=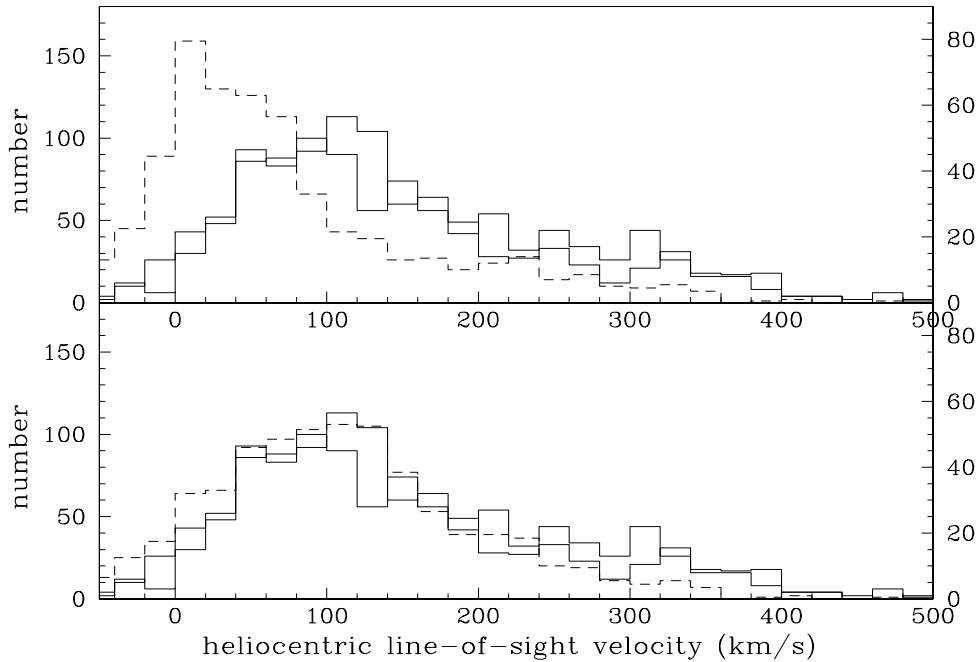,height=3.7in,width=5.2in,angle=270}
\caption{Modified from Gilmore, Wyse \& Norris 2002. In each panel the
solid histograms are observational data for faint $V \simlt 19.5$ F/G
stars in lines of sight where, at these distances, the line-of-sight
velocity probes $\sim 0.7-0.8$ of the azimuthal streaming velocity.
The dashed histogram is a model; in the upper panel the model is
derived from standard local `thick disk' kinematics which provide a
good fit to the brighter stars, while the model in the lower panel has
a significantly higher lag in $v_{rotation}$ behind the Sun.  This
provides a significantly better fit to the data.}
\end{figure}

However, the values of critical defining parameters for the
`canonical' thick disk, probed locally, remain variable from study to
study. For example, the `accepted' value for the rotational lag is
around 40km/s (e.g.~Carney, Laird \& Latham 1989), but values as low
as 20km/s (Chiba \& Beers 2000) and as high as 80km/s (Fuhrmann 2000)
have been reported.  Some of this variation is undoubtably due to the
difficulty of deconvolving a complex mix of populations.  The thin
disk will dominate any local sample, and comparison with distant {\it
in situ\/} surveys will help (cf.~the technique of Wyse \& Gilmore
1995), as will using the discrimination inherent in the distinct
elemental abundances of thick and thin disk stars (cf.~Nissen 2003).
Again, large statistically significant samples, so that tails of the
distribution functions are well-defined, in key lines-of-sight, are
needed. 

\subsection{ Small Scale Structure of the Bulge}

The bulge is clearly triaxial, but estimates of its three-dimensional
structure are hindered by dust extinction, projection effects and the
uncertainties in the structure of the disk along the line-of-sight
(e.g.~spiral arm pattern).  The inner bulge, within $\sim 1$~kpc of
the center, appears symmetric in deep infrared images taken with the
ISO satellite (van Loon et al.~2003).  The best fitting bar model
(Bissantz \& Gerhard 2002) to the COBE data has axial ratios
1:0.3--0.4:0.3 (i.e.~barely triaxial) and a length of $\sim 3.5$~kpc.
The effects of the bar potential may be the cause of the asymmetric
stellar kinematics found by Parker, Humphreys \& Beers (2003) in
samples of stars on either side of the Galactic Center. 

\subsection{Stellar Halo  Small Scale Structure}

Structure in coordinate space mixes and dissolves on dynamical
timescales.  The outer regions of the halo, say at Galactocentric
distances of greater than 15~kpc where dynamical timescales are
$\simgt 1$~Gyr, are thus most likely to host observable substructure.
Indeed, as discussed more fully in Majewski's contribution to this
volume, several streams are found in the outer halo, in both
coordinate space and kinematics.  The vast majority of the confirmed
structure is due to a single system, the Sagittarius dwarf spheroidal
(e.g.~Ibata et al.~2001; Dohm-Palmer et al.~2001; Majewski et
al.~2003; Newberg et al.~2002; Newberg et al.~2003).  This contrasts
with the predictions of many disrupted satellites in CDM models
(e.g.~Bullock et al.~2000).  The present mass of the Sagittarius dwarf
is uncertain and model-dependent, but most estimates are within a
factor of three of $10^9$M$_\odot$ (Ibata et al.~1997; Majewski et
al.~2003). The mass lost by it to the halo is also model-dependent;
presently identified streams are perhaps 15\% of the remaining bound
mass.  The evolutionary history of the Sagittarius dwarf is as yet
unclear and much work remains to be done.

Tidal streams can be, and are, also associated with dynamically
evolving globular clusters.  The excellent photometry from the Sloan
Digital Sky survey has allowed the tracing of extended, thin arms from
the outer halo globular cluster Palomar 5 over 10 degrees across the sky
(Odenkirchen et al.~2003; see Figure~3).

\begin{center}
\begin{figure}%[ht!]
\psfig{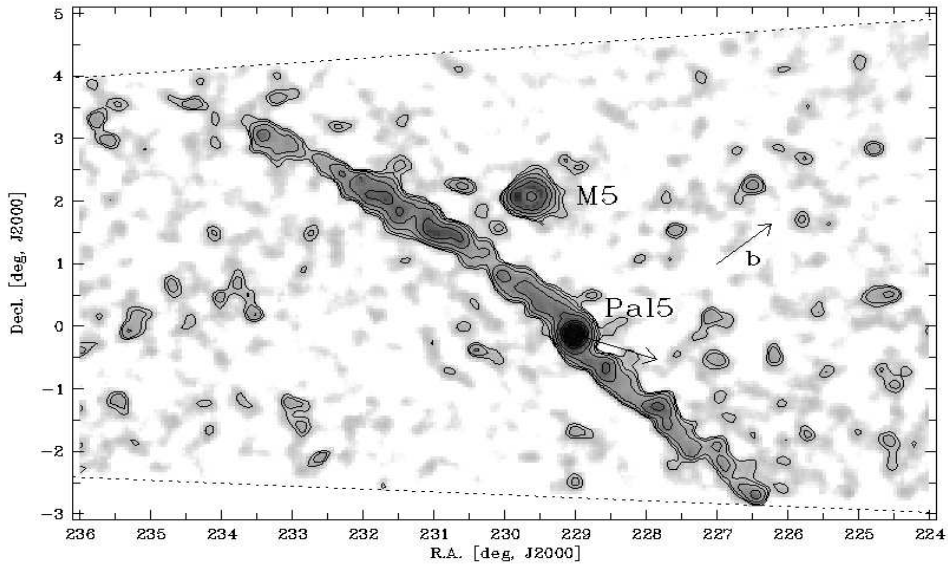}
\vskip -0.75truecm
\caption{Taken from Odenkirchen et al.~(2003), their Figure~3.  The contours show the surface density of stars that are selected from their photometry to be members of Pal~5. There are clearly  streams associated with this globular cluster.  The arrow extending from the core of Pal~5 indicates the estimated direction of its orbit.}
\end{figure}
\end{center}
Streams are rare in the inner halo (which contains most of the stellar
mass!).  Simulations suggest that signatures in phase space,
particularly if integrals of the motion can be estimated, can survive
for $\sim$ a Hubble time. A moving group has indeed be isolated (Helmi
et al.~1999), but its mass is uncertain (see Chiba \& Beers 2000), as
is its origin -- perhaps even it is associated with the Sagittarius
dwarf (e.g.~Majewski et al.~2003).

No structure is seen in coordinate space of the inner halo (within a few kpc of the Sun); the 2pt
correlation function for main sequence stars brighter than $V=19$ is
flat (Gilmore, Reid \& Hewett 1985; Lemon et al.~2003).  This rules
out significant recent accretion events that penetrate into the inner
Galaxy, and ongoing disruption of inner globular clusters.  Other tests for
substructure show low-significance features consistent with known
streams from the Sagittarius dwarf (Lemon et al.~2003), in agreement 
with results from blue horizontal branch stars (Sirko et al.~2003). 

\section{Concluding Remarks}

The properties of the stellar populations of the Milky Way contain
much information about the star formation history and mass assembly
history of the Galaxy.  The Milky Way has merged with, is merging
with, and will merge with, companion galaxies, which contribute stars,
gas and dark matter.  Debris from the Sagittarius dwarf galaxy
dominates recent accretion into the outer Galaxy, while the data are consistent with 
little stellar accretion into the inner Galaxy, including the
disk. Predominantly gaseous accretion is relatively unconstrained, and
is favoured by models of chemical evolution (cf.~Tosi's contribution).
Planned and ongoing large spectroscopic surveys will tightly constrain
the existence and orgins of stellar phase-space substructure.  The relatively
quiescent merging history of the Milky Way that is implied by the
mean properties of the stellar components is rather atypical in
$\Lambda$CDM cosmologies.  What about the rest of the Local Group?

\end{document}